\documentclass[12pt]{article}  
\usepackage{lscape}
\usepackage[dvips]{graphicx}
\begin{document}
\title{Electrosorption of Br and Cl on Ag(100): \\
Experiments and Computer Simulations\footnote{We dedicate 
this paper to the memory of Michael~J.\ Weaver. 
Mike's breadth, depth, and productivity, 
as well as his engaging personality made him an outstanding member of 
our community. We will miss him deeply.}}
\author{
\centerline{I. Abou Hamad$^{1,2}$, Th. Wandlowski$^{3}$, G. Brown$^{2,4}$, 
and P.A. Rikvold$^{1,2,5}$}\\
\centerline{\scriptsize\it $^{1}$Center for Materials Research and Technology 
and Department of Physics, Florida State University, 
Tallahassee, FL 32306-4350, USA}\\
\centerline{\scriptsize\it $^{2}$School of Computational Science and Information
Technology, Florida State University, Tallahassee, Florida 32306-4120, USA}\\
\centerline{\scriptsize\it $^{3}$Institute for Thin Films and Interfaces, ISG 3,
J\"{u}lich GmbH, D-52425 J\"{u}lich, Germany}\\
\centerline{\scriptsize\it $^{4}$Center for Computational Sciences, Oak Ridge
National Laboratory, Oak Ridge, TN 37831, USA}\\
\centerline{\scriptsize\it $^{5}$Center for Stochastic Processes in Science and 
Engineering, Department of Physics,}\\
\centerline{\scriptsize\it Virginia Polytechnic Institute and State 
University, Blacksburg, VA 24061-0435, USA}
}
\maketitle
\begin{abstract}
We present chronocoulometry experiments and equilibrium Monte Carlo
simulations for the electrosorption of Br and Cl on Ag(100)
single-crystal electrode surfaces. Two different methods are used to
calculate the long-range part of the adsorbate-adsorbate
interactions. The first method is a truncated-sum approach, while the
second is a mean-field-enhanced truncated-sum approach. To compare the
two methods, the resulting isotherms are fit to experimental
adsorption isotherms, assuming both a constant electrosorption valency
$\gamma$ and also a coverage-dependent $\gamma$. While a constant
$\gamma$ fits the Br/Ag(100) well, a coverage-dependent or
potential-dependent $\gamma$ is needed for Cl/Ag(100).
\end{abstract}

\section{Introduction}
\label{sec:I}

Specifically adsorbed anions strongly influence the structure and
dynamics of adsorbed layers on electrode surfaces~\cite{Magn,Wand02}.
Among these systems, halides electrosorbed on single-crystal metal
surfaces have been extensively studied both experimentally and
theoretically over the last decade. Examples of these studies include
the electrosorption of Cl, Br, I, and F on low-index surfaces of
Ag~\cite{VALETTE:AG,OCKO:BR/AG,Wand01} employing classical
electrochemical techniques~\cite{VALETTE:AG,Wand01,A.Hamelin,B.M.Jovic},
(see also literature cited in ref.~\cite{Wand01}), and structure sensitive
techniques such as electroreflectance~\cite{C.Franke}, {\it in-situ\/}
scanning tunneling microscopy (STM)~\cite{G.Aloisi,T.Yamata}, surface
X-ray scattering (SXS)~\cite{OCKO:BR/AG,Wand01,B.M.Ocko}, and X-ray
absorption fine structure~(XAFS)~\cite{ENDO99}.

Ocko, Wang, and Wandlowski~\cite{OCKO:BR/AG} showed that the sharp
peak observed in cyclic voltamograms of Br electrosorption on Ag(100)
corresponds to a continuous phase transition in the layer of adsorbed
Br.  This transition separates a low-coverage disordered phase at more
negative electrode potentials from a c$\left( 2 \times 2 \right) $
ordered phase at more positive potentials.  Moreover, in recent static
and dynamic studies, Mitchell, Brown, and Rikvold~\cite{A,B} have
numerically investigated the phase ordering and disordering mechanisms
in cyclic-voltammetry (CV) and sudden potential-step
experiments. Using Monte Carlo simulations of a lattice-gas model
first used by Koper~\cite{KOPER:HALIDE,KOPE98C}, they also produced
theoretical adsorption isotherms for this system that were fit to
experimental adsorption isotherms. From these fits, parameters such as
the electrosorption valency, $ \gamma $, and the next-nearest neighbor
interaction parameter, $\phi_{\rm nnn}$, were estimated.

Less attention has been given to Cl electrosorption on Ag(100),
perhaps because previous studies \cite{VALETTE:AG} have assumed that
this system is similar to Br/Ag(100). In the present paper we compare
Monte Carlo simulation results with experimental equilibrium isotherms
obtained by chronocoulometry for both Cl and Br, using a lattice-gas
model and two different methods for calculating the long-range part of
the configuration energies.

The remainder of this paper is organized as follows.  The experimental
procedures and results are discussed in Sec.~\ref{sec:Experimental}.
The lattice-gas model is presented in Sec.~\ref{sec:LGM}. Equilibrium
simulations using the two different methods of calculation are
discussed in Sec.~\ref{sec:Methods}, and the fitting procedures and
resulting parameter estimates for each system are discussed in
Sec.~\ref{sec:Fitting}.  Finally, a summary and conclusions are
presented in Sec.~\ref{sec:Conclusions}.

\section{Experimental}
\label{sec:Experimental}
The electrochemical measurements were performed with Ag(100)
single-crystal electrodes (4~mm diameter and 4~mm thickness) using the
so-called hanging meniscus technique.  The silver electrodes were
chemically etched in cyanide solution before each experiment until
the surface appeared shiny~\cite{A.Bewick} and, after careful rinsing
in Milli-Q water, annealed in a hydrogen flame for about 30~s. After
cooling in a stream of argon the electrode was quickly transferred
into the electrochemical cell. Contact with the electrolyte was
established under strict potential control, usually at values close to
the potential of zero charge of $0.05$~M $\rm{KClO}_{4}$ ($E_{\rm pzc}
\approx -0.900$~V)~\cite{M.L.Foresti}.  The counter electrode was a
platinum wire, and a saturated calomel electrode (SCE) in a side
compartment served as reference. All potentials are quoted with
respect to the SCE. The temperature was $(20 \pm 1)^{\circ}$C.

The set-up and procedures for the electrochemical experiments (cyclic
voltammetry, capacitance measurements, chronocoulometry) were
described previously~\cite{Wand01,T.Pajkossy}. We focus here only on
some additional details of the chronocoulometric
experiments~\cite{Th.Wandlowski}: The potential was initially held at
a value $E_{\rm i}$ between $-1.375$~V and $-0.300$~V ($25$~mV
interval length), up to $300$~s for the lowest chloride
concentrations, and then stepped to the final potential $E_{\rm f}=
-1.400$~V, where chloride is completely desorbed from the electrode
surface. The waiting time at $E_{\rm i}$ was chosen (in control
experiments) to be always sufficient to establish adsorption
equilibrium.  A comprehensive account of the setup and detailed
results of the SXS experiments for Ag(100)/Cl will be given
elsewhere~\cite{Th.Wandlowski}.

\subsection*{Results of the electrochemical experiments}
 
\begin{figure}
\centerline{\includegraphics[angle=0,width=3.8in]{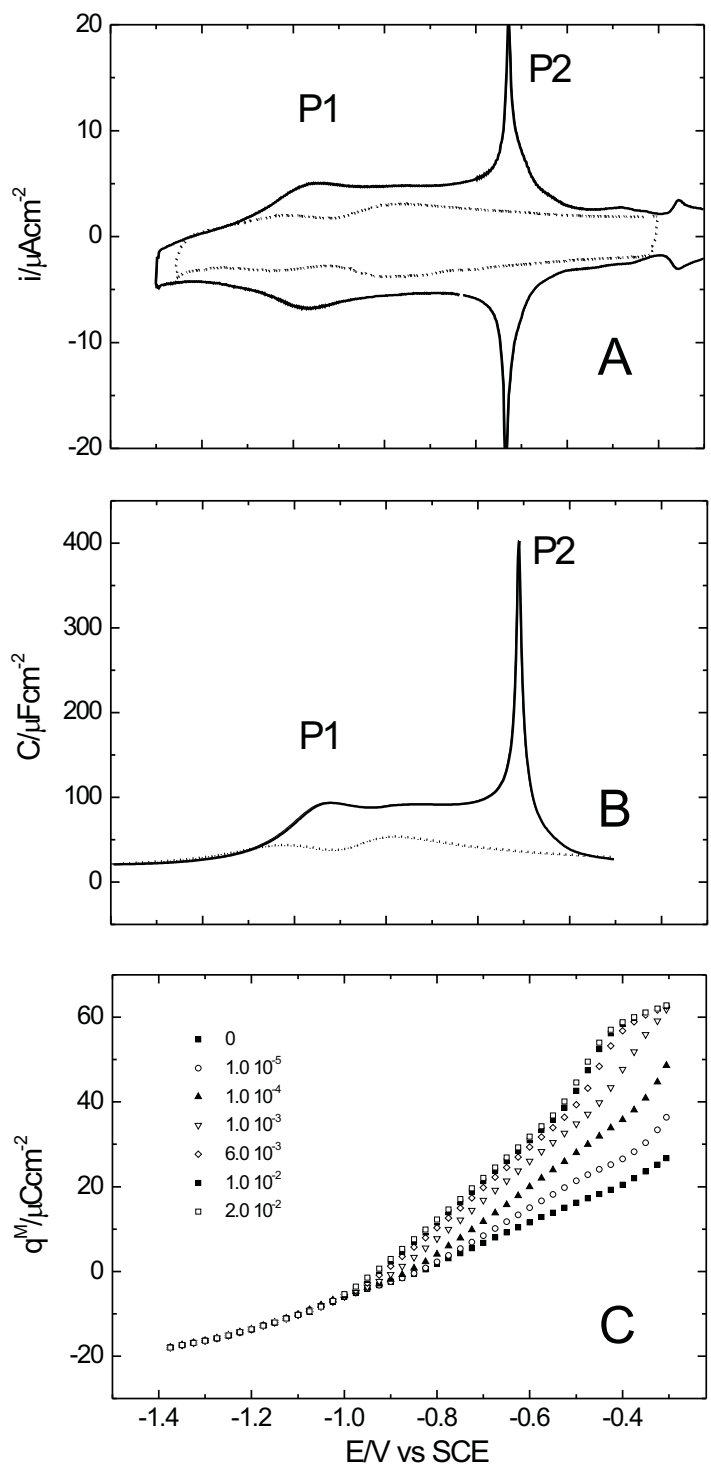}}
\caption[]{(A) Cyclic voltamogram (50~mV $\rm s^{-1}$) for
Ag(100)/$\left( 0.05-x \right) \rm{M~KClO}_{4} + {\mathit x}~\rm{M~KCl}$,
$x$=0 (dotted curve) and $x$=0.02~M (solid curve). 
(B) Capacitance vs.\ potential
curves for the system of Fig.~\ref{fig:W1}(A), 10~mV~$\rm s^{-1}$.
(C) Charge density vs.\ potential curves for
Ag(100)/$\left( 0.05-x \right) \rm{M~KClO}_{4} + {\mathit x}~\rm{M~KCl}$
($x$ in M), as indicated in the figure. 
\label{fig:W1}}
\end{figure}

Figures~\ref{fig:W1}(A) and~\ref{fig:W1}(B) show voltammograms (CV) and capacitance
(CE) curves ($5$~mV peak-to-peak amplitude, $18$~Hz ac-frequency) for
the Ag(100) electrode in $0.05$~M $\rm{KClO}_{4}$ and $0.02
\rm{~M~}KCl + 0.03 \rm{~M~KClO}_{4}$ as obtained after continuous
cycling of the potential between $-1.40$~V and $-0.10$~V.  The
capacitance curves merge at $E<-1.20$~V, indicating complete
desorption of chloride. The onset of chloride adsorption is
characterized by broad peaks in the CV and CE curves, which are
la belled P1 in Fig.~\ref{fig:W1}. {\it In-situ\/} SXS experiments
revealed that chloride adsorbs in this potential region as a lattice
gas corresponding to randomly adsorbed species on fourfold hollow
sites~\cite{Th.Wandlowski}. The sharp peak P2 in Figs.~\ref{fig:W1}(A)
and~\ref{fig:W1}(B) represents the continuous transformation of the lattice gas
into an ordered c$\left( 2 \times 2 \right) $ chloride
adlayer~\cite{Th.Wandlowski}. This phase is stable until rather
complex ${\rm AgCl}_{x}$-surface compounds are being formed at more
positive potentials [$E > -0.20$~V for the system shown in
Fig.~\ref{fig:W1}(A)]~\cite{B.M.Jovic,Th.Wandlowski}.

Equilibrium data on the adsorption of chloride ions at Ag(100)
electrodes were obtained from chronocoulometric measurements in a
gently stirred solution. Figure~\ref{fig:W1}(A) shows selected charge
density curves for KCl in $\left( 0.05-x \right) {\rm M~KClO}_{4} +
x~\rm{M~KCl}$. Adsorption of chloride causes a positive charge to flow
to the metal side of the inter­face. This is accompanied by a negative
shift of the potential of zero charge. Phenomenologically, the charge
density vs.\ potential curves as measured in the presence of chloride
concentrations $\rm{C}_{\rm Cl} > 10^{-3}$ M are composed of two
segments. The region of low charge densities corresponds to the
potential range of the broad features in the CV and CE curves. The
second one appears at a potential just positive of the adlayer phase
transition at P2, corresponding to charge densities between 35 and 40
$\mu \rm{C~cm}^{-2}$. It is characterized by an initial increase of
$\delta {q}^{\rm M}/ \delta E$. Subsequently the slope decreases and
seems to become less dependent on the chloride concentration for
$\rm{C}_{\rm Cl} > 10^{-3}$~M at the most positive potentials studied.
 
\begin{figure}
\includegraphics[angle=0,width=6in]{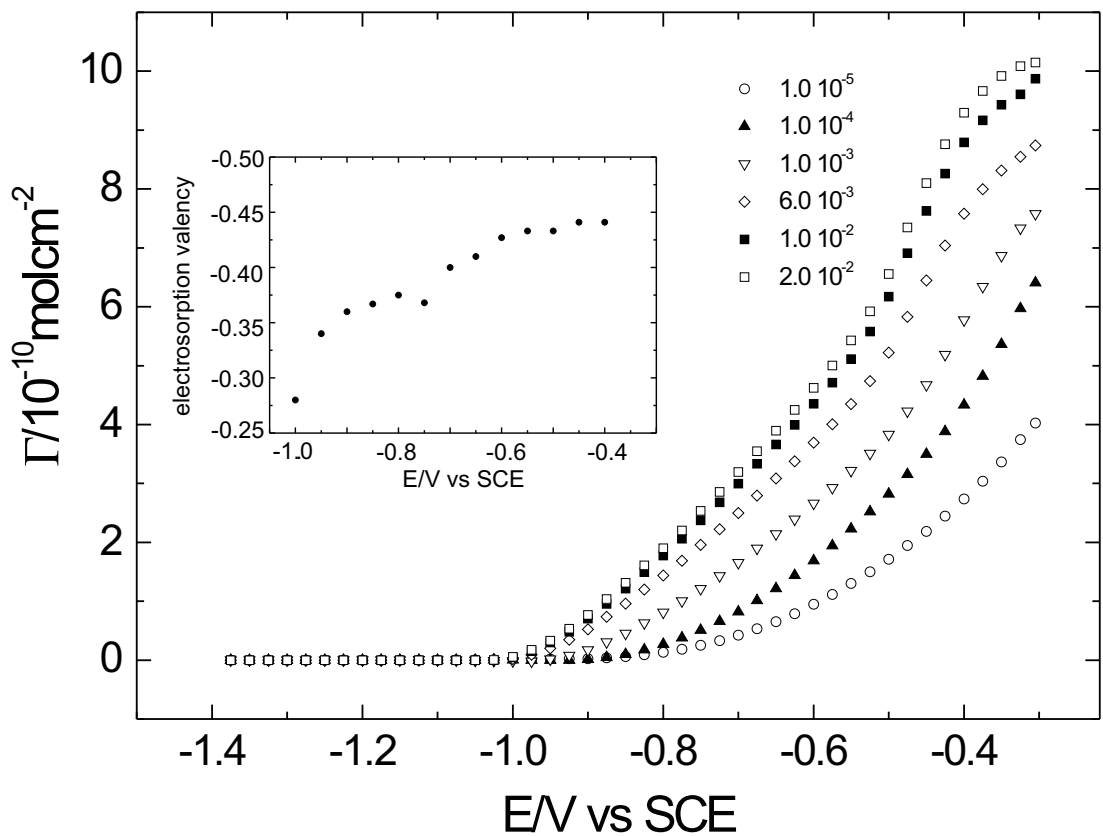}
\caption[]{Plots for surface excess for
Ag(100)/$\left( 0.05-x \right) \rm{M~KClO}_{4} + {\mathit x}~\rm{M~KCl}$
($x$ in M), employing the electrode potential as the independent electrical
variable for selected chloride concentrations. The inset shows the
potential-dependent electrosorption valency.
\label{fig:W2}}
\end{figure}

The corresponding Gibbs surface excess has been calculated at
constant potential (see Fig.~\ref{fig:W2}) and constant charge (not
shown, c.f.~\cite{Th.Wandlowski}), using the approach of Stolberg and
Lipkowski~\cite{Wand01,Th.Wandlowski,L.Stolberg}. The surface excess
increases in the potential region of disordered chloride and at
constant concentration monotonously with potential. At $\rm{C}_{\rm
Cl} > 10^{-3}$~M one observes in the potential region of the sharp
volumetric peak P2 a distinct increase in slope ($\Gamma \sim 6
\times 10^{-10} \rm{~mol~cm}^{-2}$). At the most positive potentials
experimentally accessible, the surface excess levels off and
approaches a chloride-concentration independent limit of $\Gamma_{\rm
m} = 10.2 \times 10^{-10} \rm{~mol~cm}^{-2}$. This chronocoulometric
result is close to $9.94 \times 10^{-10} \rm{~mol~cm}^{-2}$ expected
for a complete $\rm c \left( 2 \times 2 \right)$ chloride adlayer, and
is well supported by {\it in-situ\/} SXS experiments employing the
so-called interference scattering technique~\cite{Th.Wandlowski}.

The electrosorption valency, as determined from a plot of the charge
density vs.\ surface excess at constant potential for the Ag(100)/Cl
system is shown as an inset in Fig.~\ref{fig:W2}. Using this method,
the electrosorption valency is found to be potential dependent,
decreases with increasing potential, and approaches -0.45 at
-0.40~V. This value is larger than the result for Ag(100)/Br ($\gamma
\sim -0.80$) \cite{Wand01,A} (see also discussion in
Sec.~\ref{sec:Fitting}), indicating the stronger ionic character of
adsorbed chloride, in comparison to bromide on silver.

\section{Lattice-gas Model}
\label{sec:LGM}

\begin{figure}	
\includegraphics[angle=0,width=6in]{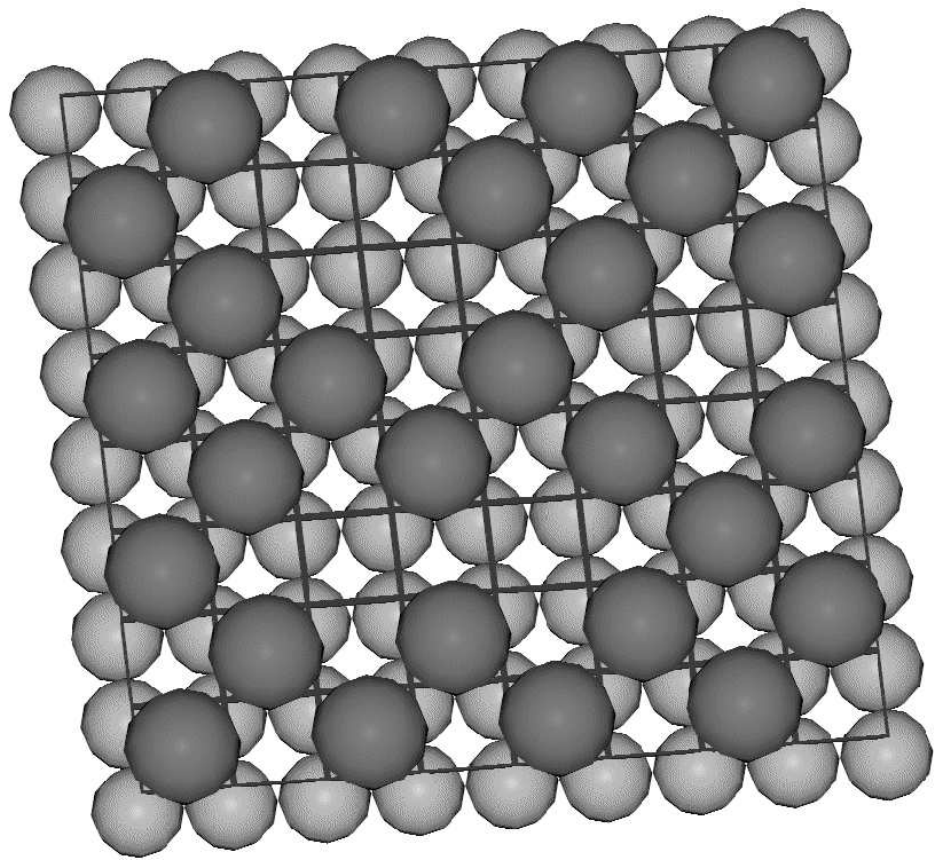}
\caption[]{Br or Cl (bigger spheres) adsorbed at the 4-fold hollow
sites of the (100) surface of Ag (smaller spheres).
The squares of the grid frame correspond to the adsorption sites.
\label{fig:model}}
\end{figure}

The model used for both the Br and Cl systems is a lattice-gas model
similar to that used by Koper~\cite{KOPER:HALIDE,KOPE98C} and Mitchell
\textit{et al.}~\cite{A,B}. In this model the Br or Cl ions adsorb at
the four-fold hollow sites of the Ag(100) surface, as shown in
Fig.~\ref{fig:model}. The model is defined by the grand-canonical
effective Hamiltonian
\begin{equation}
{\mathcal{H}} = - \sum_{i<j} \phi_{ij} c_{i} c_{j} - \overline{\mu} 
\sum_{i=1}^{L^2}c_{i}
\label{eq:H}
\end{equation}
where $\sum_{i<j}$ is a sum over all pairs of sites, $ \phi_{ij} $ are
the lateral interaction energies between particles on the $i$th and
$j$th sites measured in meV/pair, and $ \overline{ \mu } $ is the
electrochemical potential measured in meV/atom. Here the local
occupation variables $ c_{i} $ can take the values 0 or 1, depending
on whether site $ i $ is occupied by an ion (1) or solvated (0).

The simulations were performed on an $L\times L$ square lattice, using
periodic boundary conditions to reduce finite-size effects.  The
interaction constants $ \phi_{ij} $ between ions on sites $i$ and $j$
a distance $r_{ij}$ apart (measured in Ag(100) lattice spacing units,
$a=2.889$~{\AA} \cite{OCKO:BR/AG}) are given by
\begin{equation}
\phi_{ij} = \left\{ \begin{array}{lc}
                        - \infty & r_{ij}=1 \\
                        \frac{2^{3/2} \phi_{\rm nnn} }{r_{ij}^{3}} 
                                 & r_{ij} \geq \sqrt{2}
\end{array} \right.
\end{equation}
where the infinite value for $r_{ij}=1$ indicates nearest-neighbor
exclusion, the interaction for large $r_{ij}$ is most likely a
combination of dipole-dipole and surface-mediated elastic interactions,
and negative values of $\phi_{ij}$ denote repulsion. The
previous studies by Koper~\cite{KOPER:HALIDE,KOPE98C} have shown that
large but finite nearest-neighbor repulsion has only minor effects on
the coverage isotherms.  Consequently, we have chosen a simplified
model with nearest-neighbor exclusion.

In the weak-solution approximation, the electrochemical potential $
\overline{ \mu } $ is related to the bulk ionic concentration $C$ and
the electrode potential $E$ (measured in mV) as
\begin{equation} 
\overline{\mu} = \overline{\mu}_{0} + k_{\mathrm B} T \ln \frac{C}{C_{0}} 
- e \gamma E
\label{eq:mbar}
\end{equation}
where $ \overline{\mu}_{0} $ is an arbitrary constant, $C_{0}$ is a
reference concentration (here taken to be $1~\rm mM$), $ e $ is the
elementary charge unit, $ \gamma $ is the electrosorption
valency~\cite{C}, and $ \overline{ \mu } $ has the sign convention
that $ \overline{ \mu } > 0$ favors adsorption.

The anion coverage is defined as
\begin{equation}
\theta = N^{-1} \displaystyle \sum_{i=1}^{N}{c_{i}}~,
\end{equation}
where $N=L^2$ is the total number of lattice sites. The coverage can
be experimentally obtained by standard electrochemical methods as well
as from the integer-order peaks in surface X-ray scattering (SXS)
data~\cite{OCKO:BR/AG,Th.Wandlowski}.

To compute the isotherms $\theta ( \overline{\mu} ) $ or $\theta (E)$,
the change in energy must be calculated before each attempted Monte
Carlo move. This requires evaluation of the interaction sum in
Eq.~(\ref{eq:H}), which is computationally intensive even for modest
system sizes. In the next section we compare a direct summation
approach with a more efficient truncated summation combined with a
mean-field approximation for the influence of adparticle pairs at
large separations.

\section{Monte Carlo Methods}
\label{sec:Methods}
The Monte Carlo simulation proceeds as follows. On a square lattice
with $L=64$ and periodic boundary conditions we randomly select a
lattice site, $i$, and attempt to change the occupation variable
($c_{i}=1 \rightarrow c_{i}=0$ or $ c_{i}=0 \rightarrow c_{i}=1$).
The energy difference, $\Delta \mathcal{H}$, between the intial state
of the system and the proposed state is then calculated using
Eq.~(\ref{eq:H}). We use the Metropolis acceptance
probability~\cite{Land00}
\begin{equation}
\mathcal{R} = \min \left[ 1, 
\exp \left(-\frac{\Delta \mathcal{H}}{k_{\rm B}T} \right) \right]
\label{eq:P}
\end{equation}
for accepting the new trial configuration as the next configuration in
the Monte Carlo sequence.

Since the value of $\Delta \mathcal{H}$ determines $\mathcal{R}$ in
Eq.~(\ref{eq:P}), the approximations made to calculate the
large-$r_{ij}$ contributions to the pair sum in Eq.~(\ref{eq:H}) are
important. In our computations, we used two different
methods to calculate $\Delta \mathcal{H}$. The first was a
direct-summation-plus-truncation method previously used by Mitchell
\textit{et al.}~\cite{A,B}. In the second method, which we introduce
here, we include the exact contribution of fewer sites on the surface,
using a mean-field approximation to calculate the contribution from
all the sites at larger distances.

\subsection{Direct summation}
\label{sec:nomean}

\begin{figure}
\includegraphics[angle=0,width=6in]{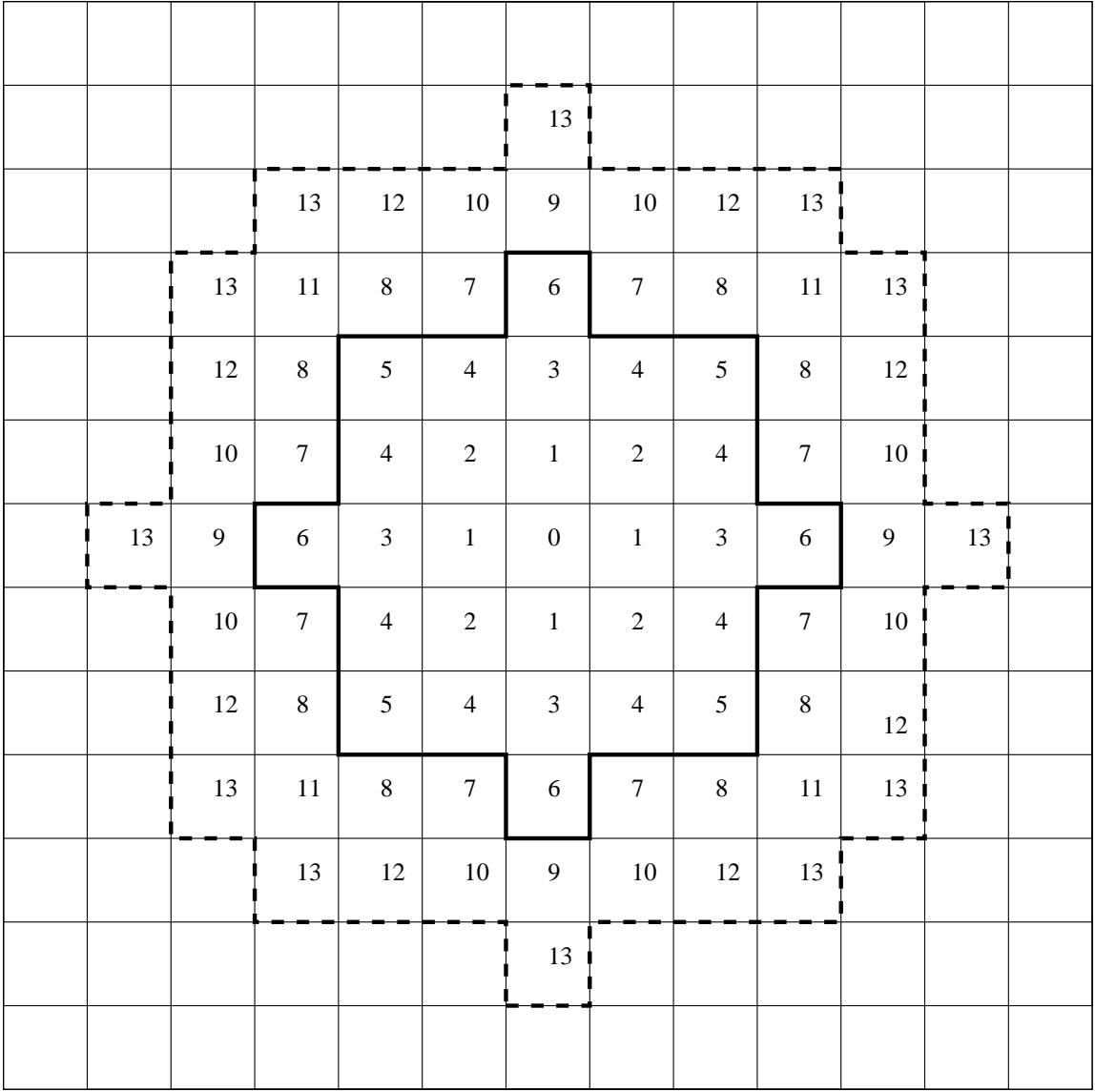}
\caption[]{Sites included in the energy calculation. 
The dashed lines surround the 
76 sites included in the direct summation for $1 < r \le 5$. 
The heavy solid lines enclose the 24 sites included in the direct 
sumation for $1 < r \le 3$, which is used with the mean-field-enhanced method.
\label{fig:sites}}
\end{figure}

The first method involves calculating the contribution of the sites
that are less than or equal to five lattice spacings away from the
site of interest.  Since nearest-neighbor adsorption is prohibited,
the acceptance probability vanishes for any move that would bring two
particles into nearest-neighbor positions. In addition to the four
nearest-neighbor sites, there are a total of 76 sites such that $1 <
r_{ij} \le 5$. These are divided between 12 different distance classes
as shown in Fig.~\ref{fig:sites}. This summation provides
approximately 87\% of the total energy due to the long-range
interactions~\cite{B}.

This method of approximation proved to be acceptable for simulating Br
adsorption on Ag(100)~\cite{A,B} as it produced values for the
electrosorption valency $\gamma$ that are in good agreement with
experimental data. On the other hand, because of its computationally
intensive nature, this method of calculating $\Delta \mathcal{H}$ made
it harder for dynamic simulations to achieve CV scan rates within the
experimental range~\cite{B}.

\subsection{Mean-field-enhanced method}
\label{sec:mean}
The method of calculating $\Delta \mathcal{H}$ can be modified from
the previous one so that, without reducing the accuracy of the
calculation, the cutoff radius is reduced from five to three lattice
constants.  This decreases the number of sites beyond nearest
neighbors that are included in the summation from 76 to 24, as shown
in Fig.~\ref{fig:sites}.

This modification includes a mean-field approximation to estimate the
interaction energy contribution from those particles that are farther
than three lattice constants from the site in question. Consequently,
the change in energy associated with desorbing from the surface a
particle at lattice site $i$ is
\begin{equation}
\Delta \mathcal{H}_i = 2^{3/2} \phi_{\rm nnn} \left[ \sum_{1 < r_{ij} \leq 3} 
\frac{c_{j}}{r_{ij}^3} +
\theta \left( \Sigma_\infty - \sum_{1 \leq r_{ij}\leq 3} 
                             \frac{1}{r_{ij}^3} \right) \right]
\label{eq:deltaE}
\end{equation}
which is both more efficient and more accurate than calculating the
exact energy contributions out to $r_{ij}=5$.  Here $\Sigma_\infty =
\sum_{r=1}^{ \infty} 1/r^3$ is a Zucker sum over the sites of a square
lattice of unit lattice constant \cite{KOPER:HALIDE}, and the quantity
in the parenthesis is a constant, independent of the adsorbate
configuration. It can easily be calculated to arbitrary numerical
precision. The energy change upon adsorption at a previously empty
site $i$ with four empty nearest-neighbor sites is the negative of the
result in Eq.~(\ref{eq:deltaE}).

This modification of the calculation of the energy changes for the
Monte Carlo steps makes the program run more than twice as fast as
when using the method described in Sec.~\ref{sec:nomean} and used in
Refs.~\cite{A,B}.  It is consequently better suited for dynamical
simulations \cite{ABOU03}.

The results of the calculations performed with this method are similar
to those using exact summation for $r\leq 5$. At higher coverages, the
numerical effect of introducing the mean-field approximation is more
pronounced, as expected. We note that this method is successful
because the interactions in our model are repulsive, which ensures
that the spatial distribution of adparticles is approximately uniform
on large length scales.  For systems with attractive interactions more
sophisticated methods to calculate the contributions from the
long-range interactions, such as the Fast Fourier Transform or Fast
Multipole Method \cite{FMM}, would be needed.

\subsection{Comparison}
\label{sec:compare}
\begin{figure}
\begin{center}
\includegraphics[angle=-90,width=6in]{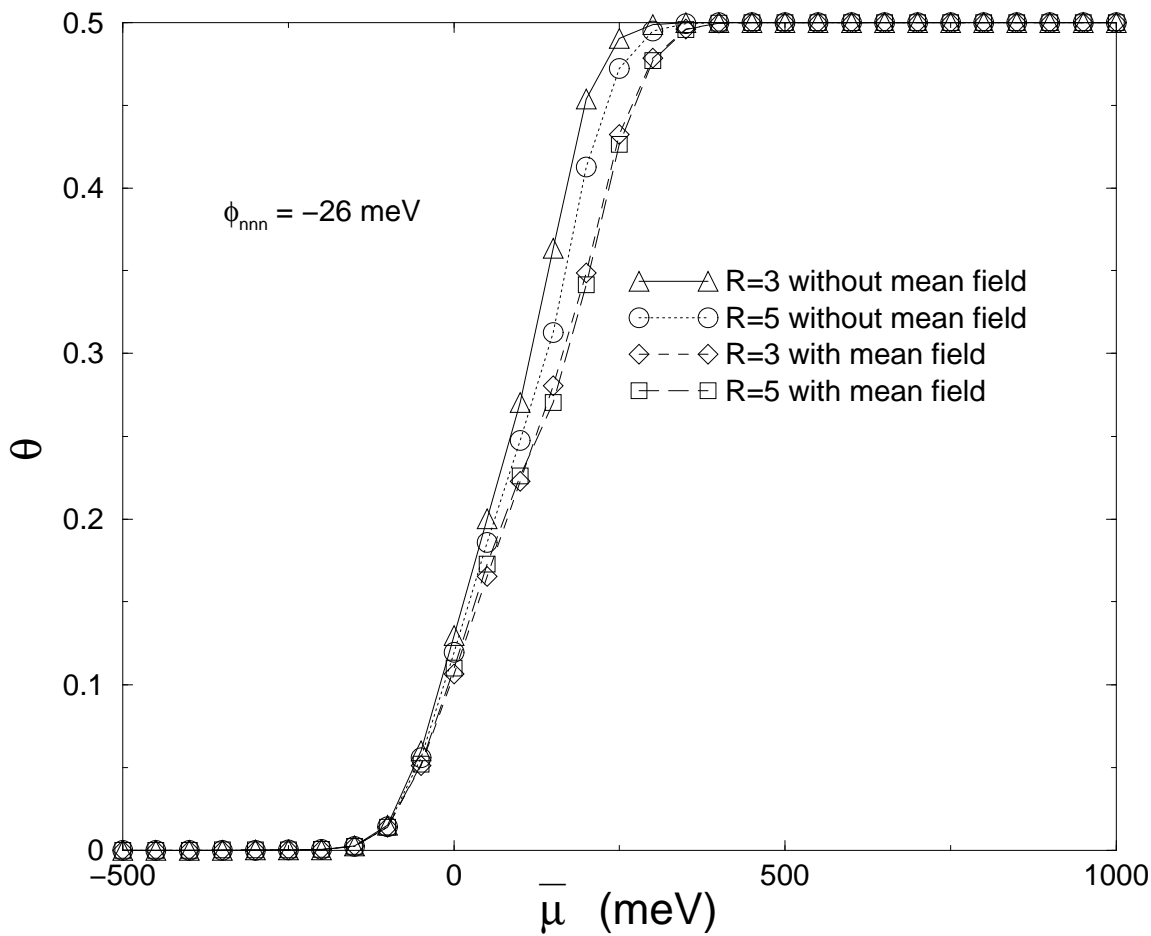}
\caption[]{Simulated coverage isotherms vs.\ electrochemical 
potential, computed with different methods. Here $R$ is
the cutoff radius up to which exact summation was performed.
The simulations were performed at $T \approx 27^\circ$C with 
$\phi_{\rm nnn} = - 26$~meV. The line segments connecting the data
points are included as
a guide to the eye. 
\label{fig:meannomean}}
\end{center}
\end{figure}

A comparison of the calculated coverage as a function of the
electrochemical potential, obtained with both mean-field and
non-mean-field methods for different cutoff radii, is shown
in~Fig.~\ref{fig:meannomean}.  The mean-field enhancement makes a
significant contribution to the adsorption isotherms for both cutoff
lengths. However, when the mean-field enhancement is used, increasing
the cutoff radius beyond three does not significantly improve the
results. Tests with even smaller cutoff radii revealed that three is
an optimal choice to increase the computational speed with only a
minimal loss of accuracy.

\section{Parameter Estimation}
\label{sec:Fitting}

To estimate the parameters in the lattice-gas model, $\phi_{\rm nnn}$
and $\gamma$, we performed standard equilibrium MC simulations
\cite{Land00} with $\Delta \mathcal{H}$ calculated by the
mean-field-enhanced method at room temperature ($k_{\rm B} T=25$~meV
or $T\approx290$~K) to obtain $\theta(\overline{\mu})$ for different
parameter values. These simulated isotherms were then compared with
the experimental chronocoulometry data using two fitting
procedures. The first procedure assumed a constant electrosorption
valency $\gamma$, while the second assumed a coverage-dependent
$\gamma(\theta)$. The experimental coverage data for Br/Ag(100)~\cite{Wand01},
which were also used in Refs.~\cite{A,B}, were provided to the
authors of those papers by J.X.\ Wang.

\subsection{Constant $\gamma$}
\label{sec:const}
To extract the values of $ \gamma $ and $ \phi_{ \rm nnn }$ that would
best fit the experimental data, we used a least-squares fitting
procedure that minimizes $ \chi^{2} $ with respect to three fitting
parameters for Cl and Br electrosorbed on Ag(100)~\cite{B},
\begin{equation}
        \chi^{2}( \phi_{ \rm nnn } , \gamma , \overline{\mu}_{0} )
        = \sum_{k=1}^{N_{\rm conc}} \sum_{l=1}^{l_{max} (k)} 
            [ \theta^{ \rm sim } ( E_{l}^k ;
        \overline{\mu}_{0} , \gamma, \phi_{ \rm nnn } ; C_{k} ) -
        \theta^{ \rm exp } ( E_{l}^{k} ; C_{k} ) ]^{2}
\label{eq:chi}
\end{equation}
\\ where $\theta^{\rm sim}$ and $\theta^{\rm exp}$ are the simulated
and experimental coverage values respectively, $k$ is the
concentration index, $N_{\rm conc}$ is the number of different
concentrations used in a fit, and $l$ is the point index of the
experimental data points.

This fitting procedure, which provides good fits for Br/Ag(100) in
Refs.~\cite{A,B} (see also Table~1) does not give satisfactory fits
for Cl/Ag(100). The assumption that the electrosorption valency $
\gamma $ is independent of the coverage is consistent with the Br
results, but appears to be invalid for the Cl system.

\subsection{$\theta$ dependent $\gamma$}
\label{sec:var}
\begin{figure}
\begin{center}
\includegraphics[angle=-90,width=4.5in]{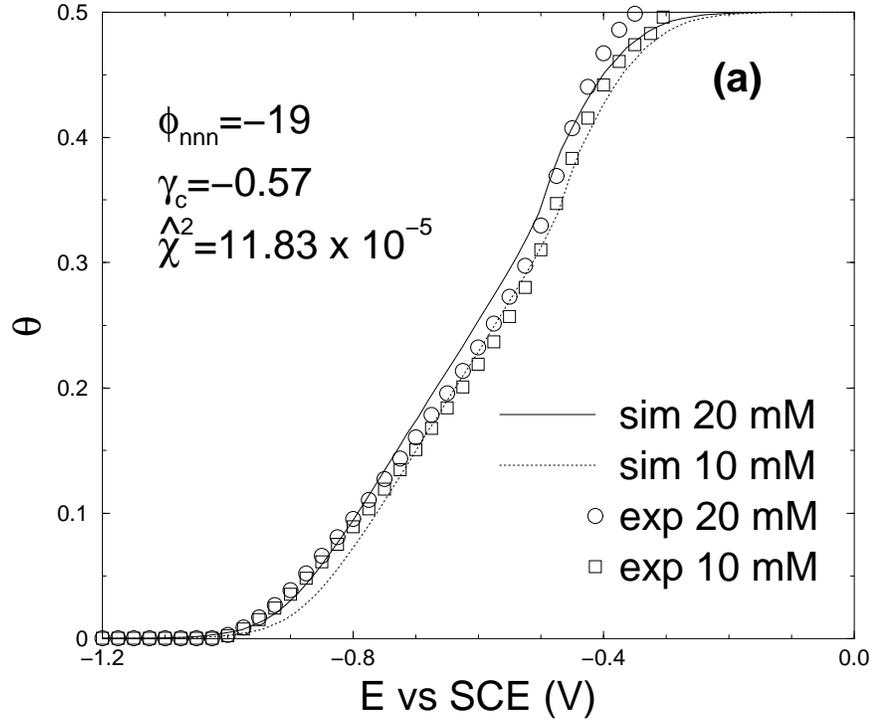}
\includegraphics[angle=-90,width=4.5in]{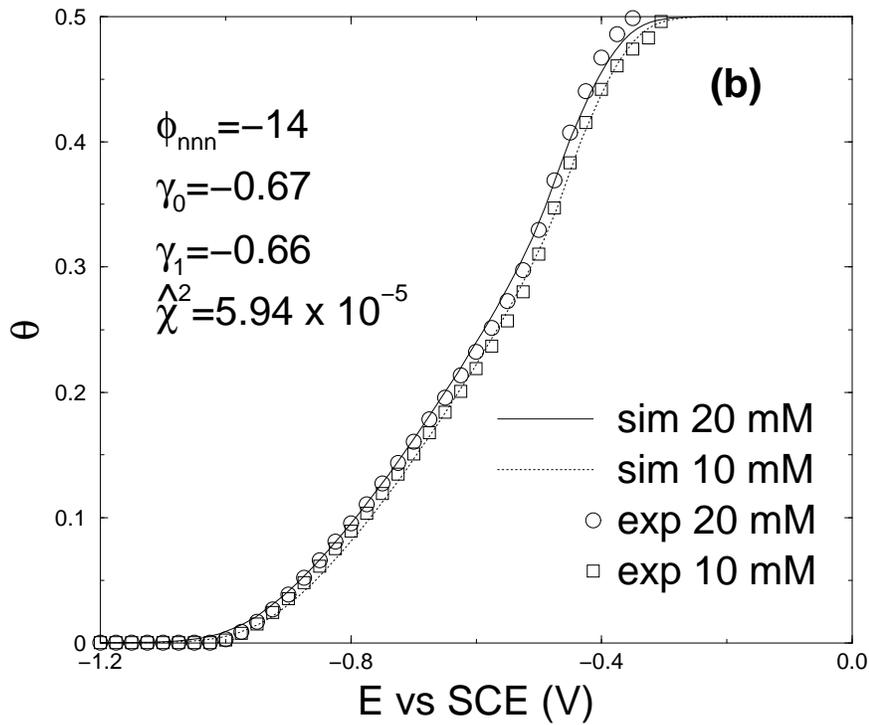}
\vspace{.2in} 
\caption[]{One simulated Cl/Ag(100) isotherm fit to two experimental isotherms
  with different bulk concentrations. The simulated isotherms were
obtained using the mean-field-enhanced
method described in Sec.~4.2.   (a) Using coverage independent
  $\gamma$.  (b) Using coverage-dependent $\gamma ( \theta)=
  \gamma_{0} + \gamma_{1} \theta$. \label{fig:fit2}}
\end{center}
\end{figure}

Since a coverage independent $ \gamma $ does not appear to be a valid
assumption for Cl/Ag(100), we assume that $\gamma$ depends linearly on
the coverage as
\begin{equation}
        \gamma (\theta) = \gamma_{0} + \gamma_{1} \theta~
\label{eq:gamma}
\end{equation}
This adds to the model a fourth parameter, $\gamma_1$, that must be
determined by comparison to the experimental results.  The
$\theta$-dependent $\gamma$ is allowed to vary in the interval
$[-1,0]$, which limits the values that $\gamma_{0}$ and $\gamma_{1}$
can assume according to Eq.~(\ref{eq:gamma}). The range of variation
for $\gamma_{0}$ is $[-1,0]$, while it is $[-2,0]$ for $\gamma_{1}$
since the latter is multiplied by $\theta$, which can assume a
maximum value of 0.5.  As determined by $\chi^2$
[Eq.~(\ref{eq:chi})], the inclusion of a coverage-dependent
electrosorption valency improves the fits to the adsorption isotherms,
shown in Fig.~\ref{fig:fit2}.  It also gives reasonable values for $
\gamma $ and $ \phi_{ \rm nnn } $. The resulting parameter values,
based on fitting Cl and Br simulations to the experimental isotherms,
are shown in Tables~1 and~2, respectively.

The tables include values for the fitting parameters, $\gamma_{\rm c}$
for the constant $\gamma$ fits and $\gamma_{0}$ and $\gamma_{1}$ for
the coverage-dependent fits. Moreover they include values for
$\phi_{\rm nnn}$ and $\chi^{2}$ per degree of freedom, $\hat{\chi}^2$,
for all the fits.  Fitting to the experimental data was performed both
for the truncated-sum Monte Carlo isotherms and for the
mean-field-enhanced isotherms. For each of these simulated isotherms
we performed three kinds of fits. First, the numerical isotherms were
fit to a single set of experimental data with a certain halide ion
concentration~(columns 3-8 in Tables~1 and~2, labeled 1-fit). Second,
each numerical isotherm was simultaneously fit to two experimental
isotherms with ionic concentrations of 10~mM and 20~mM for Cl or 1~mM
and 10~mM for Br (columns 9 and 10, labeled 2-fit). Third, numerical
isotherms were fit to three experimental isotherms simultaneously with
the concentrations 6, 10 and 20~mM for Cl and 0.1, 1.0, and 10.0~mM
for Br (columns 11 and 12, labeled 3-fit).

In fitting two or more isotherms simultaneously, several conclusions
may be drawn from the tables shown. With one exception, the values of
$\gamma_{1}$ in Table~1 for Br/Ag(100) are zero (i.e.,
$|{\gamma_{1}}|<0.01$) or nearly zero for all of the fits. This
supports the conclusion that a constant $\gamma$ is a good
approximation for that system.  Moreover, the quality of the fits is
enhanced by using two or more concentrations simultaneously because
this utilizes a greater number of experimental data points in one fit
and constrains $\gamma$ more directly through the $\ln(C/C_{0})$ term
of Eq.~(\ref{eq:mbar}). For constant $\gamma$ the 2-fit and 3-fit
values for $\phi_{\rm nnn}$ and $\gamma$ for the mean-field-enhanced
method are consistent with those obtained by the non-mean-field
method. However, smaller values of $\hat{\chi}^2$ are found for
the mean-field-enhanced method than for the non-mean-field method,
indicating that the mean-field-enhanced isotherms fit slightly better
to the experimental data than the non-mean-field ones. Comparing the
values found using the 2-fit and 3-fit methods for mean-field-enhanced
isotherms, suggests $\phi_{\rm nnn}=-21\pm 2$ meV and $\gamma_{\rm
c}=-0.71\pm 0.01$ as reasonable estimates. These results are
consistent with those obtained in Refs.~\cite{Wand01,A,B}.
  
The results for the fits to the Cl/Ag(100) experiments are given in
Table~2, whose layout is the same as for Table~1.  Notice that
$\gamma_{1}$ assumes non-zero values for most of the fits, and that
-- except~for~the~fit~to~the~6~mM~isotherm -- the values of $\hat{\chi}^2$
obtained with variable $\gamma$ are less than those obtained with the
constant-$\gamma$ fits. This supports the conclusion that, for the Cl
system, a coverage-dependent $\gamma$ should be adopted. The exception
can be attributed to the fact that the 6 mM experimental data points
do not cover the whole $\theta$ range, especially in the important
region near the phase transition.  Consequently, we hereafter ignore the 6~mM
data in estimating parameters for the Cl/Ag(100) system. This leaves
the 2-fit parameter values for a $\theta$-dependent $\gamma$. The
mean-field and non-mean-field parameters are similar and, as
determined by $\hat{\chi}^2$, provide equally valid descriptions of
the experiment. From this we conclude that, for Cl/Ag(100), $\phi_{\rm
nnn}=-14\pm 2$ meV and $\gamma = (-0.66\pm 0.01)+(-0.68\pm
0.01)\theta$.  While these results for $\gamma_0$ and $\gamma_1$ are
more negative than those obtained from the charge-density curves in
Sec.~\ref{sec:Experimental}, the trends of a stronger dependence on
coverage or electrode potential and a higher (less negative) overall
value than for Br/Ag(100) are the same. The reasons for the quantitative
difference between the estimates based on the two methods to determine
the electrosorption valency are left for future study.

\section{Conclusions}
\label{sec:Conclusions}

In this paper we have presented a new method for implementing the
lattice-gas model for the electrosorption of halide ions on a Ag(100)
surface by introducing a mean-field-enhanced method to calculate the
contributions to the configuration energy of widely separated
adparticle pairs. We fit adsorption isotherms obtained by Monte Carlo
simulation to the experimental adsorption isotherms based on
chronocoulometric data after comparing them to the results obtained
using the non-mean field method.

The isotherms from both methods were fit to experimental data for both
Cl and Br. It was shown that it was safe to assume a
coverage-independent or potential-dependent electrosorption valency
for Br on Ag(100) while a coverage-dependent electrosorption valency
had to be implemented for the Cl on Ag(100).

\section*{Acknowledgments}
We thank S.J.\ Mitchell, J.X.\ Wang, and B.M.\ Ocko for useful
discussions. P.A.R.\ thanks the
Physics Department of Virginia Polytechnic Institute and State
University for hospitality during the final stages of this work.

Supported by US National Science Foundation Grant No.\ DMR-9981815 and
by Florida State University through the Center for Materials Research
and Technology and the School of Computational Science and Information
Technology.

%\bibliography{full.bib}{10}

%\bibliography{/home/scri42/users/rikvold/decstation/tex/biblio/elchem,refs.bib}
%\bibliographystyle{unsrt}

\clearpage 

\begin{landscape}
\begin{table}
\caption[]{Fitting parameters for Br adsorption on a Ag(100) single-crystal 
surface, using isotherms computed with a cutoff distance of three lattice 
constants, and with and without mean-field interactions included. 
The quantity $ \hat{\chi}^2 $ is the 
$\chi^{2}$ per degree of freedom of the fit. 
Here, ``1-fit" means individual 
fits for each of the three different bulk concentrations given; 
``2-fit" means a single, simultaneous fit for the two bulk concentrations 
given; and 
``3-fit" means a single, simultaneous fit for the three bulk concentrations 
given. 
}
\begin{footnotesize}
\begin{tabular}{|c|c|c|c|c|c|c|c|c|c|c|c|}
\hline
\multicolumn{2}{|c|}{} 
  & \multicolumn{6}{|c|}{1-fit; 0.1 mM, 1 mM, 10 mM} 
    & \multicolumn{2}{|c|}{2-fit; 1 mM, 10 mM} 
      & \multicolumn{2}{|c|}{3-fit; 0.1mM, 1 mM 10, mM} \\ 
\cline{3-12}
\multicolumn{2}{|c|}{}
  & \multicolumn{3}{|c|}{mean-field} 
    & \multicolumn{3}{|c|}{no mean-field} 
      & mean-field & no mean-field & mean-field & no mean-field \\ 
\hline	
  & $\gamma_{\rm c}$
    &$-0.51$&$-0.49$&$-0.46$
      &$-0.47$&$-0.46$&$-0.45$
        &$-0.70$&$-0.67$
          &$-0.73$&$-0.70$ \\ 
\cline{2-12}
$\gamma=\rm{Const.}$ 
  & $\phi_{\rm nnn}$ (\rm{meV})
    &$-9$&$-10$&$-10$
      &$-8$&$-10$&$-11$
        &$-22$&$-23$
          &$-22$&$-23$ \\ 
\cline{2-12}
  & $\hat{\chi}^2\times10^5$ 
    &$0.564803$&$2.14327$&$7.34073$
      &$0.680078$&$2.56305$&$8.09803$
        &$6.89425$&$9.46991$		
          &$7.31550$&$9.76821$ \\ 	
\hline	
  &$\gamma_{0}$	
    &$-0.54$&$-0.49$&$-0.46$	
      &$-0.53$&$-0.46$&$-0.45$	
        &$-0.70$&$-0.66$	
          &$-0.72$&$-0.68$ \\ 	
\cline{2-12}	
  &$\gamma_{1}$					
    &$-0.01$&$0.00$&$0.00$	
      &$0.02$&$0.00$&$0.00$
        &$0.00$&$0.04$
          &$-0.01$&$-0.19$ \\ 	
\cline{2-12}	
$\gamma=\gamma_{0}+\gamma_{1}\theta$ 	
  & $\phi_{\rm nnn (meV)}$		
    &$-10$&$-10$&$-10$		
      &$-11$&$-10$&$-11$	
        &$-22$&$-21$
          &$-21$&$-16$ \\ 
\cline{2-12}	
  &$\hat{\chi}^2\times10^5$	
    &$0.54580$&$2.20281$&$7.54464$	
      &$0.69541$&$2.63425$&$8.32297$	
        &$6.98496$&$9.53128$	
          &$7.36551$&$8.68241$ \\ 	
\hline	
\end{tabular}	
\end{footnotesize}	
\end{table}	
\end{landscape}	
	
\begin{landscape}	
\begin{table}	
\caption[]{Fitting parameters for Cl adsorption on a Ag(100)
single-crystal surface, using isotherms computed with a cutoff
distance of three lattice constants, and with and without mean-field
interactions included. The quantities shown and the organization of the 
table are the same as in Table~1. 
}	
\begin{footnotesize}
\begin{tabular}{|c|c|c|c|c|c|c|c|c|c|c|c|}
\hline
\multicolumn{2}{|c|}{} 
  & \multicolumn{6}{|c|}{1-fit; 6 mM, 10 mM, 20 mM} 
    & \multicolumn{2}{|c|}{2-fit; 10 mM, 20 mM}
      & \multicolumn{2}{|c|}{3-fit; 6 mM, 10 mM, 20 mM} \\
\cline{3-12}
\multicolumn{2}{|c|}{}
  & \multicolumn{3}{|c|}{mean-field}
    & \multicolumn{3}{|c|}{no mean-field}
      & mean-field & no mean-field
        & mean-field & no mean-field \\
\hline
& $\gamma_{\rm c}$ 
  &$-0.29$&$-0.35$&$-0.40$
    &$-0.28$&$-0.35$&$-0.46$
      &$-0.57$&$-0.54$
        &$-0.44$&$-0.44$ \\
\cline{2-12}
$\gamma=\rm{Const.}$
  & $\phi_{\rm nnn}$ (\rm{meV})
    &$-4$&$-6$&$-8$
      &$-4$&$-7$&$-14$
        &$-19$&$-20$
          &$-12$&$-14$ \\ 
\cline{2-12}
  & $\hat{\chi}^2\times 10^5$
    &$4.06844$&$5.05880$&$11.2260$
      &$4.11751$&$4.63378$&$10.3544$
        &$11.8302$&$11.7546$
          &$13.2892$&$13.1223$ \\
\hline
&$\gamma_{0}$
  &$-0.29$&$-0.46$&$-0.66$
    &$-0.28$&$-0.47$&$-0.65$
      &$-0.67$&$-0.65$
        &$-0.43$&$-0.42$ \\
\cline{2-12}
&$\gamma_{1}$
  &$-0.00$&$-0.36$&$-0.68$
    &$-0.00$&$-0.36$&$-0.67$
      &$-0.66$&$-0.70$
        &$0.02$&$0.04$ \\
\cline{2-12} $\gamma=\gamma_{0}+\gamma_{1}\theta$
& $\phi_{\rm nnn}$ (\rm{meV}) 
  &$-4$&$-7$&$-12$
    &$-4$&$-9$&$-14$
      &$-14$&$-14$
        &$-11$&$-12$ \\
\cline{2-12}
  &$\hat{\chi}^2\times10^5$
    &$4.17015$&$3.35255$&$5.61327$
      &$4.22045$&$2.74380$&$4.28522$
        &$5.93654$&$5.41895$
          &$13.2942$&$13.1225$ \\
\hline
\end{tabular}
\end{footnotesize}
\end{table}
\end{landscape}

\end{document}